\documentclass{egpubl}
\usepackage{EGUKCGVC2016}

 \WsPaper         
 \electronicVersion 

\ifpdf \usepackage[pdftex]{graphicx} \pdfcompresslevel=9
\else \usepackage[dvips]{graphicx} \fi

\PrintedOrElectronic

\usepackage{t1enc,dfadobe}

\usepackage{egweblnk}
\usepackage{cite}

\usepackage{amsmath}
	
\title[Topological Visualisation techniques for the understanding of Lattice Quantum Chromodynamics simulations]%
      {Topological Visualisation techniques for the understanding of Lattice Quantum Chromodynamics (LQCD) simulations}

\author[D.P. Thomas, R. Borgo and S. Hands]
       {D.P. Thomas$^{1,2}$, R. Borgo$^{1}$
        and S. Hands$^{2}$
        \\
         $^1$Department of Computer Science, College of Science, Swansea University, United Kingdom\\
         $^2$Department of Physics, College of Science, Swansea University, United Kingdom
       }

\graphicspath{{./images/}}

\begin{document}


\maketitle

\begin{abstract}
   The use of topology for visualisation applications has become increasingly popular due to its ability to summarise data at a high level.  Criticalities in scalar field data are used by visualisation methods such as the Reeb graph and contour trees to present topological structure in simple graph based formats.  These techniques can be used to segment the input field, recognising the boundaries between multiple objects, allowing whole contour meshes to be seeded as separate objects.  In this paper we demonstrate the use of topology based techniques when applied to theoretical physics data generated from Quantum Chromodynamics simulations, which due to its structure complicates their use.  We also discuss how the output of algorithms involved in topological visualisation can be used by physicists to further their understanding of Quantum Chromodynamics.
   \\


\end{abstract}

\section{Introduction}

Scientific visualisation often relies upon indirect volume rendering as a method for translating discrete data into visual representations for analysis.  The core part of this process is in creating an accurate model of the data with a particular set of parameters.  The Marching Cubes algorithm~\cite{lorensen1987marching} presented a significant step forward in computing objects as three dimensional surfaces that could be rendered and interacted with as objects in 3D space.

Whilst presenting an initial method for visualising data in a new way, a number of improvements and optimisations have been made since the introduction of Marching Cubes.  A key step forward was the ability to segment data into connected components, referred to as isocontours, allowing discrete objects to be seeded from a single sample point in the data~\cite{VanKreveld1997}.  Topology led methods, such as the contour tree~\cite{tarasov1998construction}, provide compact data structures to encapsulate and represent isocontours of scalar data as the function value varies.  Topological changes in the data are captured as vertices in the tree structure.

This paper discusses how the contour tree and a direct relation, the Reeb graph, can be used to visualise data from a branch of theoretical physics named Quantum Chromodynamics.  Beyond allowing us to efficiently compute meshes for visualisation, we show how the structure of the data can also be captured and presented using topological visualisation techniques.  Finally, we summarise how we believe this can help to push forward the boundaries of understanding in Quantum Chromodynamics.

The remainder of this paper is structured as follows, in section~\ref{sec::qcd} we introduce relevant concepts of Quantum Chromodynamics (QCD) and review existing uses of volume visualisation in Quantum Chromodynamics.  Section~\ref{sec::applications} introduces relevant visualisation techniques by giving examples of how they were applied to the QCD data.  We present key contributions that we believe are of use to QCD scientists in section~\ref{sec::contributions}, gathered in part from domain expert feedback as detailed in section~\ref{sec::case_studies}.  Finally, in section~\ref{sec::conclusion} we reflect upon how topological visualisation techniques can extend the set of tools available to domain specialists and introduce future avenues of research.

\section{Lattice Quantum Chromodynamics}
\label{sec::qcd}
Quantum Chromodynamics is the modern theory of the strong interaction between elementary particles called quarks and gluons, responsible for binding them into strongly-bound composites called hadrons. The most familiar examples are the protons and neutrons found in atomic nuclei.  The most systematic way of calculating the strong interactions of QCD is a computational approach known as lattice gauge theory or lattice QCD. Space-time is discretised so that field variables are formulated
on a four-dimensional hypercubic lattice.

In table~\ref{tab::glossary} we define terms used by physicists to describe the hierarchical structure and operators used in lattice QCD. These terms are
used in the remainder of the document in the description of the domain
and work flow.

\begin{table*}[ht]
\centering
\resizebox{\textwidth}{!}{%
\begin{tabular}{ll}
Term & Definition \\ \hline

Adjoint matrix				& A complex matrix generated by transposing the input and taking the complex conjugate of the entries.\\

(Anti-)Instanton			& A four dimensional particle characterised by minima / maxima points in QCD energy fields.\\

Chemical Potential			& A parameter in lattice QCD simulations believed to induce changes in the instanton / anti-instanton balance.\\

Complex conjugate			& A complex number produced by negating the sign on the imaginary part of the input, the real part is unchanged.\\

Configuration              	& A unique lattice within an Ensemble representing the Quark-Gluon field at a specified point within the simulation.\\

Cooling                    	& Iterative algorithm used by physicists to remove noise and stabilise output from LQCD simulations.\\

Ensemble                   	& A collection of configurations created using a predefined parameter, for example chemical potential ($\mu$). \\

Euclidean space-time       	& Four-dimensional representation of Euclidean space with an additional simplified time parameter. \\

Hypercubic lattice			& A lattice that exists in four-dimensions made of hypercubic cells (see Fig. \ref{fig::lattice})\\

Periodic boundary			& A method for representing an infinite axis, where the limits (min / max) are directly connected to form a loop.\\

Plaquette                  	& A closed 2D loop around links (edges) on the lattice from an origin site.\\

Site						& Name commonly given to a vertex on the lattice represented by a co-ordinate 4-tuple ($x$, $y$, $z$, $t$). \\

Space-like plaquette       	& A plaquette existing in the XY, XZ or YZ plane (see Fig. \ref{fig::plaquettes}).\\

Time-like plaquette        	& A plaquette existing in the XT, YT or ZT plane. (see Fig. \ref{fig::plaquettes}).\\

Topological Charge Density 	& An energy field that is computed relative to each site on the lattice.\\

\end{tabular}
}
\caption{A glossary of terms specific to lattice Quantum Chromodynamics}
\label{tab::glossary}
\end{table*}

\subsection{Simulation details}
 
Lattice QCD simulations are typically generated by physicists using varying parameters in what they term an \emph{ensemble}.  This holds a number key parameters of the simulations including, but not limited to the lattice dimensions, granularity and \emph{chemical potential}.  Observables on the lattice known as \emph{instantons} and their duals, anti-instantons, are expected to be affected by chemical potential.  As this parameter is modified it is expected that the relative balance of instantons / anti-instantons will vary.  For each ensemble a number of separate \emph{configurations} are computed as part of an evolutionary Markov chain.  Generated configurations are characteristically noisy; hence, physicists use a process known as \emph{cooling} to iteratively simplify the structure of a configuration using methods respecting the underlying physics.  Physicists have established techniques for determining the correct number of cooling iterations to use in studies; hence, it is unusual to work with raw un-cooled configurations.

\begin{figure}[htb]
  \centering
  \includegraphics[width=.95\linewidth]{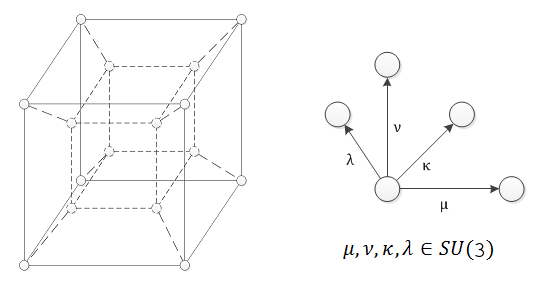}
  %
              
  \caption{\label{fig::lattice}Left: an example of a hypercubic lattice cell.  Right: link variables exist in four Euclidean dimensions.}
\end{figure}

Individual configurations are modelled using a four-dimensional hypercubic lattice.  An important implementation detail of which is the treatment of the time axis.  Unlike real-world sampled data sets the time axis is considered as an extension to conventional three-dimensional space to become four-dimensional \emph{Euclidean space-time}.  This means that physicists make no distinction between the four axes, and it is equally valid to think of three-dimensional slices of the data containing a time component as the same as an individual time-slice.  The four-dimensional lattice stores its data on the edges between vertices as \emph{link variables} (see Figure \ref{fig::lattice}).  Link variables, commonly labelled $\mu, \nu, \lambda$ and $\kappa$, are represented by a $3 \times 3$ complex matrix from the special unitary group of matrices $SU(3)$.  Physically, the link variables represent the strength of the gluon field at discrete sampling intervals.  All dimensions feature \emph{periodic boundaries}, meaning it is possible to return to an origin point by continually moving in the same direction across an axis.  The periodic nature of lattice QCD means it is possible to compute systems on relatively small lattice; typical sizes include $12^3 \times 24$, $16^3 \times 8$, where the $24$ and $8$ refer to time, but larger simulations are possible.

\begin{figure}[htb]
  \centering
  \includegraphics[width=.95\linewidth]{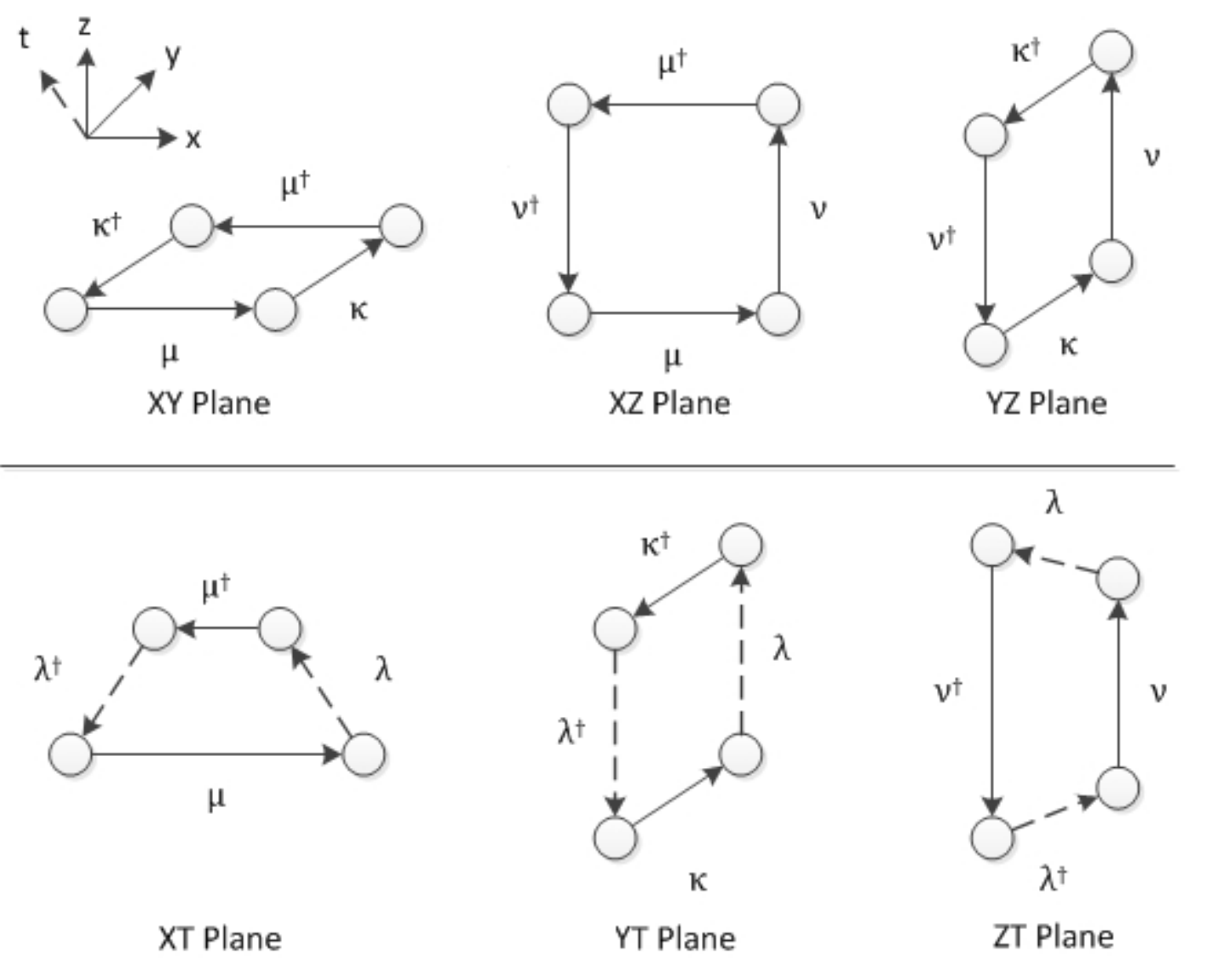}
  %
  \caption{\label{fig::plaquettes}Top: the three space-like plaquettes on the lattice.  Bottom: the three time-like plaquettes on the lattice.}
\end{figure}

\begin{figure*}[htbp]
  \centering
  \mbox{} \hfill
  \includegraphics[width=.23\linewidth]{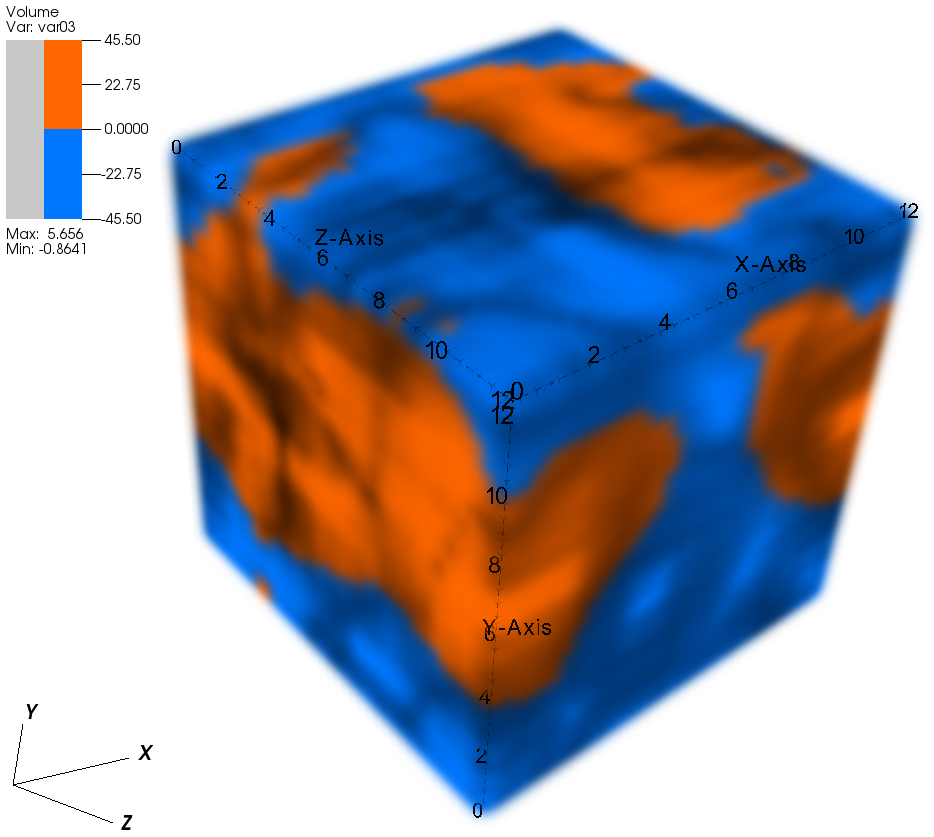}
  \hfill
  \includegraphics[width=.23\linewidth]{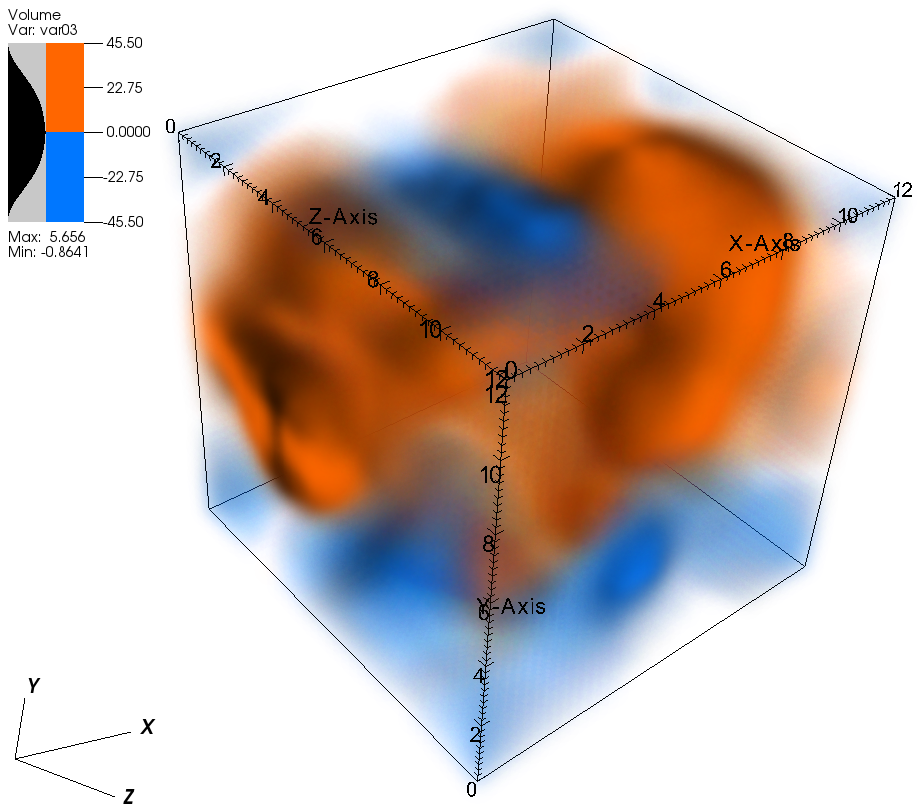}
  \hfill 
  \includegraphics[width=.23\linewidth]{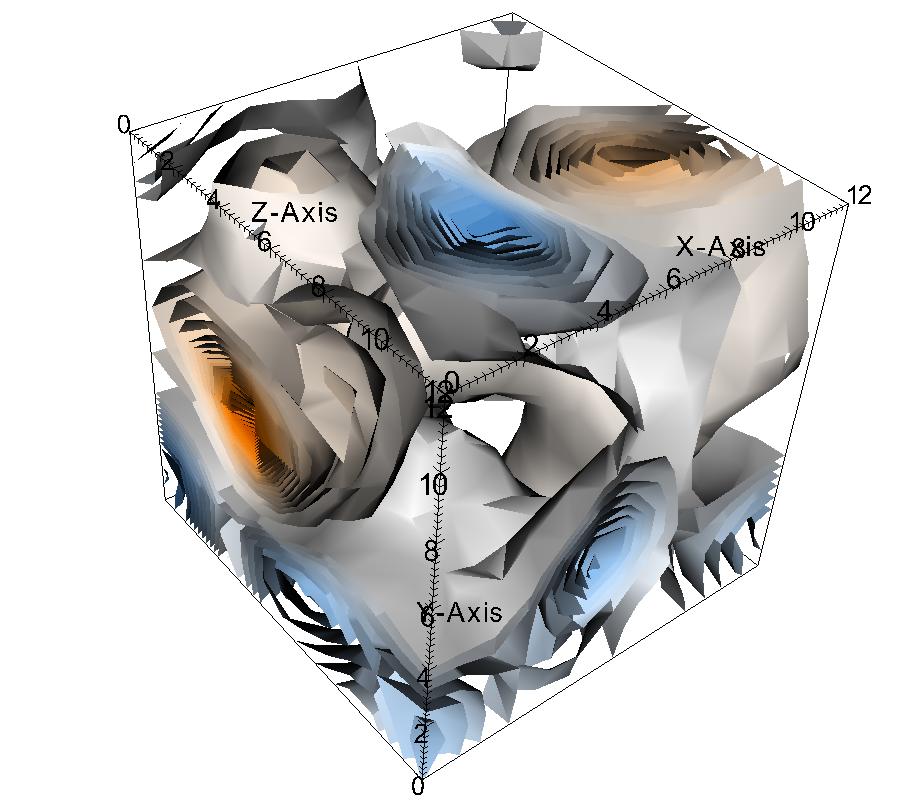}
  \hfill
  \includegraphics[width=.23\linewidth]{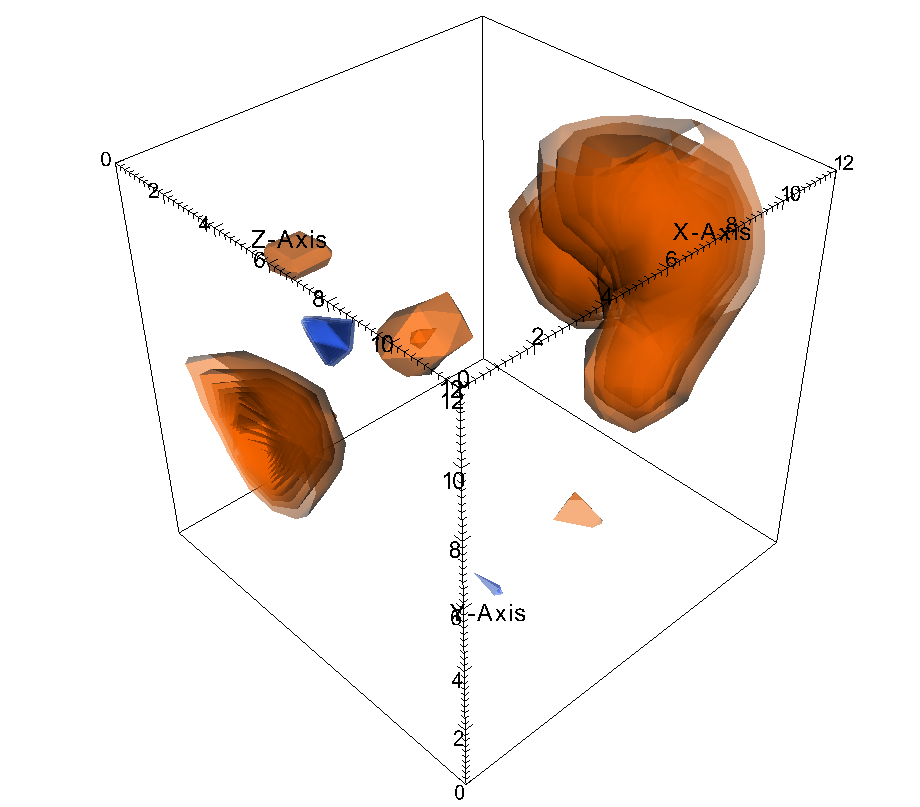}
  \hfill
  \mbox{}
  \caption{\label{fig::visit_graphics}Four views of the same data; Left-to-right: a direct volume render with no opacity transfer function, a direct volume render with a opacity transfer function tuned to remove 'noise', 100 contours computed and coloured as level sets, filtered level sets.}
\end{figure*}

\subsection{Domain specification}

In lattice QCD field strength variables are defined by navigating around the lattice. Closed loops from a given origin are referred to as a \emph{plaquette}, themselves represented by a $SU(3)$ matrix.  Plaquettes can be placed into two categories; \emph{space-like} and \emph{time-like} plaquettes (Fig. \ref{fig::plaquettes}).  Movements across the lattice in a positive direction require multiplication by the relevant link variable. Equivalently, movement in a negative direction requires multiplication by the matrix in its \emph{adjoint}, or conjugate transpose, form.  \emph{Topological charge density} is a loop around all four dimensions from an origin point, thus values are a multiplicative combination of three space-like plaquettes and three time-like plaquettes.

\subsection{Related Work}

The use of visualisation in Lattice QCD is often limited to dimension independent forms such as line graphs, used to plot properties of simulations as part of a larger ensemble.  Visualisation of a space-time nature are less frequently observed but do occasionally occur in literature on the subject.  Surface plots were used in \cite{hands1990lattice} to view 2D slices of data and give a basic understanding of energy distributions in a field. 

Application of visualisation of QCD in higher dimensions first appear in the form of three dimensional plots \cite{Feurstein1997} of instantons showing their correlation to other lattice observables.  Primarily, these were monopole loops, instantons that persist throughout the entire time domain of a simulation.  Prior to visualisation, simulations are subjected to iterative cooling to remove noise whilst keeping the desired observables intact.  Instantons are visualised using their plaquette, hypercube and L\"{u}scher definitions, each of which make it possible to view instantons and monopole loops.  Visualisation is a useful tool in this case to confirm that monopole loops and instantons are able to co-exist in a simulation, a fact that can help theoretical physicists understanding of the QCD vacuum structure.

A more detailed visualisation, making use of established computer graphics techniques, was performed by Leinweber \cite{leinweber2000visualizations} as a way of conveying to the physics community that visualisation of large data sets would help their understanding of QCD.  An area of LQCD investigated is the view that instanton-anti-instanton pairs can attract and annihilate one another during cooling phases. Topological charge density and Wilson action measurements are used to show separate spherical instantons and anti-instantons in the lattice hyper-volume.  Using animation it is possible to see that instantons in vicinity of anti-instantons deform and merge.  

More recently DiPierro et al. \cite{di2012visualization} produced several visualisations as part of a larger project with the aim of unifying various active LQCD projects.  Expectations of the work were that visualisations could be used to identify bugs within the simulation process, along with increasing understanding of LQCD.  Many existing file structures used within the LQCD community could be processed and output using the VTK visualisation library, this could then be rendered using a number of pre-existing visualisation tools including VisIt and Paraview.  In addition, the evolution of instantons over Euclidean time spans could be rendered into movies.  A web front end application and scripting interface was also provided that allowed users to set up an on-line processing chain to analyse simulations on a large scale.  Despite the apparent success of the work it would appear that it is no longer maintained and many of the links present with the project literature are inaccessible.


\section{Application of visualisation techniques to QCD data}
\label{sec::applications}

The following section describes the range of techniques that we used in gathering an understanding of the QCD structure.  Before going into specifics of how each technique was used, we provide a brief overview of the relevant topics.  Certain aspects of the data proved to be challenging when applied to commonly available techniques, such as contour tree backed visualisation, we discuss how we came about these limitations and how we addressed them in section \ref{section::reebGraphs}.

\begin{figure*}[htp]
	\centering
	\mbox{} 
	\includegraphics[width=.17\linewidth]{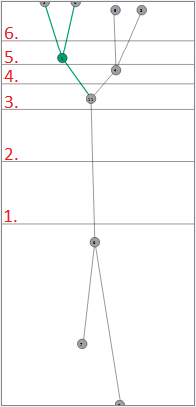}
	\hfill
	\includegraphics[width=.80\linewidth]{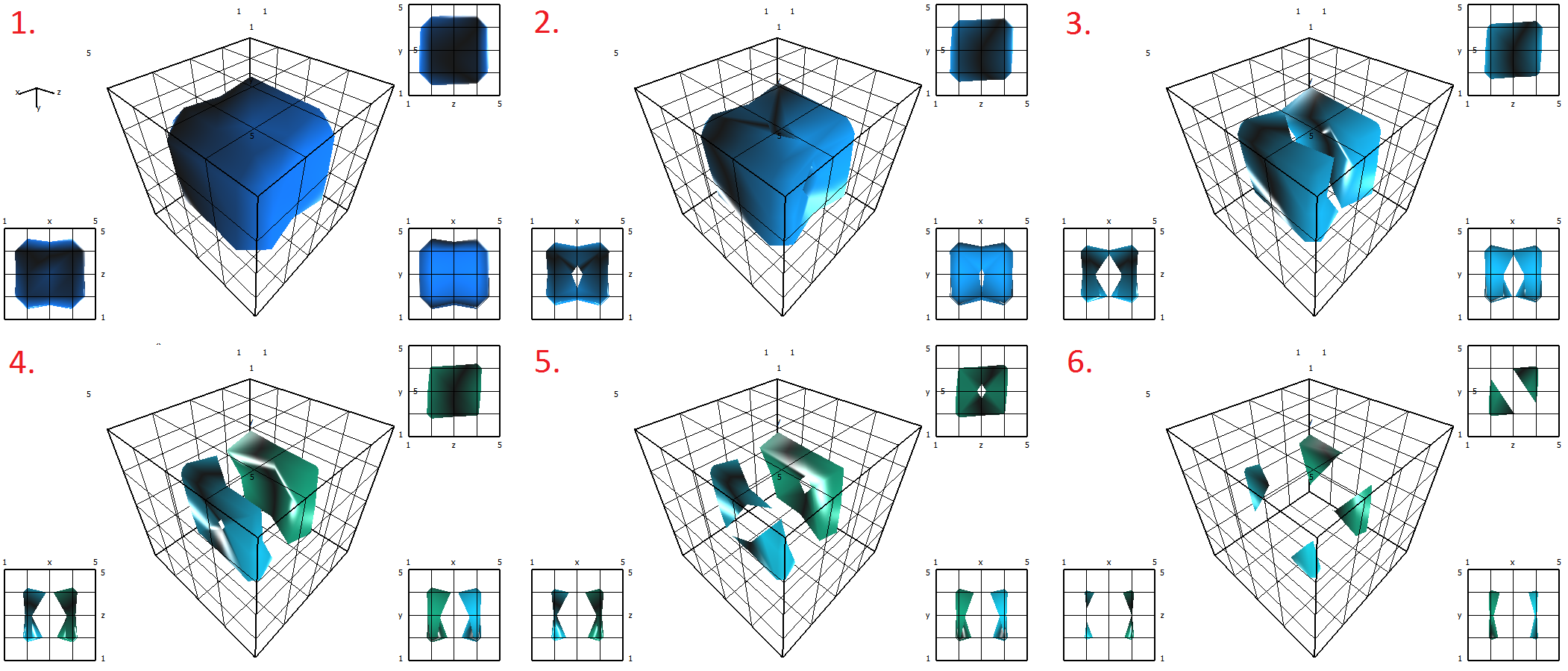}
	\mbox{}
	
	\caption{\label{fig::contour_tree_example}A contour tree representing the topology of the scalar field shown on the right.  The function height (isovalue) is represented by the height on the y-axis, contours visible at differing isovalues are characterised by the number of arcs crossing through the line representing each function height.  A large object gradually deforms from a solid to a horse shoe shaped object (1 -- 3).  The change from a topological sphere to a 2-torus at (2) is not encoded in the contour tree, as overall connectivity does not change.  At (4) the object has split into two distinct objects, represented by two arcs in the tree.  These two objects continue to deform, splitting into 3 objects at (5) and then further splitting into 4 at (6).}
\end{figure*}

\subsection{Initial approaches to visualisation}

We first used Visit~\cite{visit14} to examine our data using direct volume rendering techniques (see Fig.~\ref{fig::visit_graphics}), as it allowed us to perform a comparison with existing visualisations from similar data sources~\cite{Bennett:2013lea}.  Whilst this approach allows an overview of the data to be formed, it limits the ability to examine the structure of discrete objects within the volume due to occlusion.  Moving to indirect volume techniques, such as Marching Cubes~\cite{lorensen1987marching}, enabled visualisation of the data in a more discretised form by segmenting it into level sets.  Further refinement of the technique, through the use of multiple filters, allows viewing of nested contours within specified bounds.  Using this approach uninteresting areas of the data can be filtered out.  In lattice QCD the region centred around isovalue zero is generally considered uninteresting; using multiple filters allows the visualisation to be focused on instanton structures present at the positive/negative extremes of the data.

The existing FORTRAN code used to perform cooling of the data was able to predict the locations of minima and maxima in the Topological Charge Density field.   We used this information to locate and cross examine potential (anti-)~instantons using an isosurface representation.  Beyond predicting the positions of minima/maxima, few other properties of the objects could be revealed using the cooling tool; hence, the new  visualisations allowed domain scientists to form new questions about their structure.  One such question was if varying parameters of the physical simulation could alter the quantity and shape of objects in the volume.  This hypothesis prompted the use of topology to perform segmentation of the data into distinct objects, allowing new data querying and exploration possibilities.

\subsection{Contour Trees and flexible isosurfaces}

Contour trees, a direct derivative of Morse Theory, are able to map topological changes in a field to simple data structure (see Fig. \ref{fig::contour_tree_example}).  In \cite{VanKreveld1997a} the observation that an isosurface is the level set of a continuous function in 3D space suggested that a whole contour could be traced starting from a single element: a seed. Carr et al. \cite{Carr2003} investigated the use of contour trees in higher dimensional data-sets, whilst also improving upon the algorithm proposed in \cite{tarasov1998construction}.  They introduced the concept of augmented contour trees, an extension that added non-branching vertices to display non-critical points in the data, where non-critical vertices are used to provide values for isosurfaces seeding. Separate isosurfaces can be generated for each non-critical value identified in the contour tree, separately colour coded and manipulated by the end user.  This allows users to identify regions of interest and distinguish them accordingly \cite{carr2004simplifying}. 

\begin{figure}[htb]
  \centering
  \includegraphics[width=.95\linewidth]{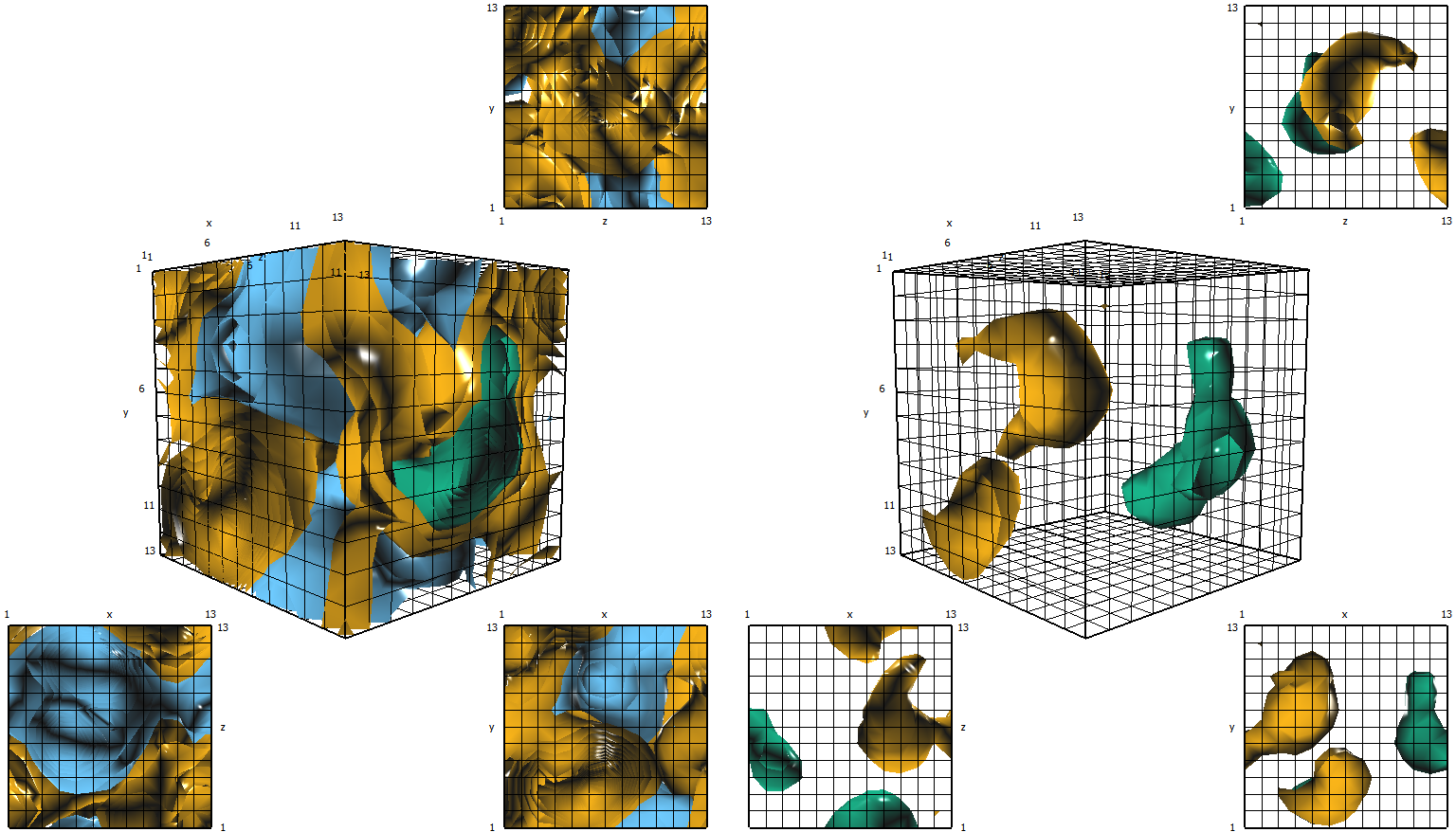}
  %
  \caption{\label{fig::surfaces}A topological charge density data set; left: shown as nested contours and right: as an individual slice at a fixed isovalue.  The green region has been selected by the user using the contour tree to highlight the same region in both views.}
\end{figure}

Recent visualisation tools have successfully used the contour tree to perform topology controlled volume rendering; hence, this was the initial method we choose to use in this work.  The contour tree captures changes within the topology of a scalar function with respect to isovalue, in this case Topological Charge Density.  As a data structure the contour tree can be considered as a dual rooted tree, covering the isovalue range, with arcs representing unique topological objects.  The algorithm works internally by combining split and join trees, collectively named merge trees, to compute \emph{critical} points in the topology.  Within the contour tree critical vertices are defined as those having degree 1, in the case of leaf nodes spawning completely new objects, or degree $\geq 2$ where the function value splits or joins an existing topological object.

We use the flexible isosurface algorithm~\cite{carr2003path} to compute and render contours at a user chosen set of isovalues.  To achieve this arcs in the tree are augmented with pointers to samples within the scalar field, allowing regions of connected components to be grouped using a union-find algorithm~\cite{tarjan1975efficiency}.  A triangle mesh is generated around a connected region, with cells at the boundary approximated using interpolated values from the surrounding scalar field.  The notion of connected regions is what allows us to move from level sets to discrete contours represented by triangular meshes.

\subsection{Reeb graphs}
\label{section::reebGraphs}

The Reeb graph is a generalisation of the contour tree capable of handling non-simply connected domains, such as those that are defined with periodic boundaries.  This presents problems to the contour tree algorithm, as it fails in the merge tree computation stage of the algorithm.  For most purposes the Reeb graph can be considered as a direct descendent of the contour tree, with the ability of allowing loops in the topology~\cite{cole2003loops}.  The first use of the Reeb graph for encoding topological features for visualisation purposes was by Shinagawa et al.~\cite{shinagawa1991constructing} where the structure was used as a way of representing objects obtained from computerized-tomography (CT) sources~\cite{shinagawa1991surface}.  

As with the contour tree, the Reeb graph can be applied to models of any dimension, provided it is represented on a simplicial mesh~\cite{pascucci2007robust}.  Tierny et al. proposed an optimized method for computing Reeb graphs using existing contour tree algorithms, by applying a procedure that they called loop surgery~\cite{tierny2009loop}.  This method was able to take advantage of the speed of contour trees with a minimal overhead in complexity.  A further optimised algorithm~\cite{doraiswamy2013computing} used the join tree of the data to identify potential loops.  An important modification upon the technique developed by Tierny et al.~\cite{tierny2009loop} was the segmentation of domain into multiple loop free contour trees, instead of a single contour tree, thus enabling multiple regions of the input to be computed in parallel.  

\begin{figure}[htb]
  \centering
  \includegraphics[width=.95\linewidth]{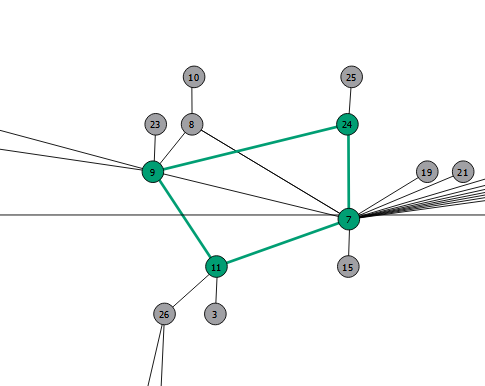}
  %
  \caption{\label{fig::reeb_loop}An example of a loop present in a Reeb graph.   Loops are common in QCD data around isovalue zero, known as the region of peculation, representing toroidal structures in the topology.}
\end{figure}

The moment calculations, described in section~\ref{section::moments}, require meshes to be a closed manifold surface, but this is often not the case in lattice QCD, due to the periodic nature of the data.  It was found that objects often span one or more boundaries, meaning the object could often be split into 2 or more distinct open objects.  The Reeb graph presents a method for solving this problem, that the contour tree could not, and has since become the default method for segmenting the data for analytical purposes.  However, the contour tree and associated flexible isosurface algorithm, still form the backbone of the rendering system where speed is of key importance.

\begin{figure*}[htb]
  \centering
  \includegraphics[width=.95\linewidth]{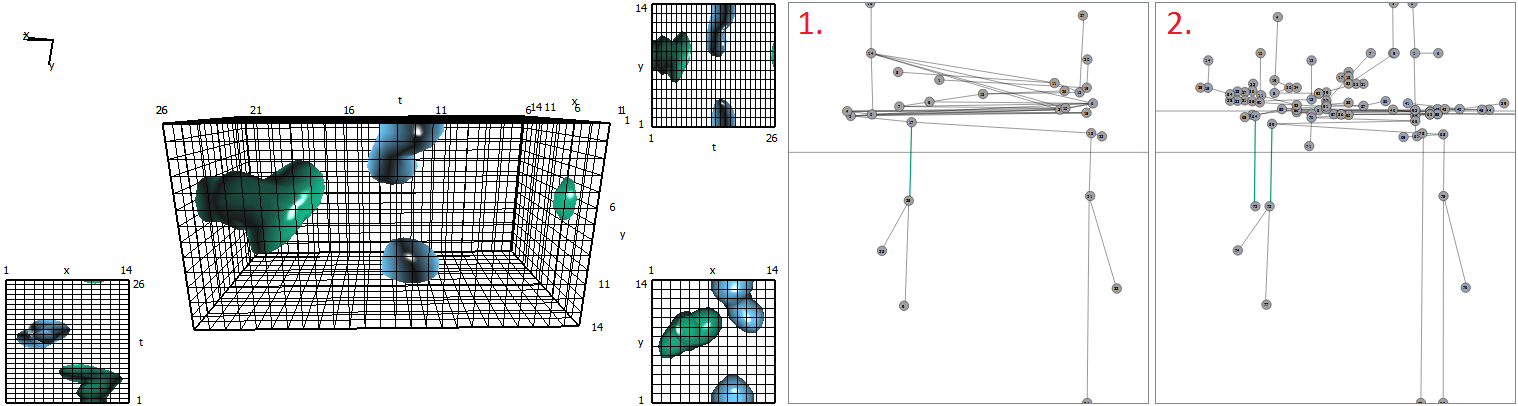}
  %
  \caption{\label{fig::periodic_boundaries}Comparison of surfaces crossing a periodic boundary on the Reeb graph (1) and the Contour Tree (2). In the left image the green object
  is crossing the periodic boundary on the time axis, whilst the blue object crosses on the y axis.  It can be seen that the green highlighted object appears twice in the contour tree (representing the two separate parts) and once in the Reeb graph.}
\end{figure*}

A direct comparison of the Reeb graph and the contour tree output is demonstrated in figure \ref{fig::periodic_boundaries}, this shows the effect that periodic boundaries have on the output of each.  Visually there is a large degree of similarity between the contour tree and the associated Reeb graph representation.  However, there are two major variations; first, the Reeb graph is no longer limited to a tree structure and can contain loops (see Fig.~\ref{fig::reeb_loop}).  Second, there is a drop in the number of vertices and arcs making up the graph due to objects spanning the domain boundary, in the contour tree these would appear as multiple arcs representing open meshed objects.  The duplication is not present in the Reeb graph, giving us a truer representation of the data from the view of the domain scientists.

\subsection{Object oriented visualisation}

The ability to compute distinct objects as features in the scalar topology opens up a number of possibilities for displaying and querying the Topological Charge Density field (Fig. \ref{fig::surfaces}).  In order to assess this, we developed a set of tools to compute properties of objects in the data, this is further discussed in section \ref{section::qcdvis}.  The ability to obtain distinct objects using the contour tree has allowed us to apply object oriented program techniques to the data, augmenting them with additional computed properties.  The concept of moments presents a generalised solution to parametrising key objects in the scalar field; these include the enclosed volume, centre-of-mass and principal component axis which is discussed in greater detail in section \ref{section::moments}.  In lattice QCD the objects that exist on the extreme branches of the contour tree, relating to the scalar field minima and maxima, are potential (anti-)instantons.  The number and physical properties of these objects are thought to be directly linked to parameters of the ensemble, particularly chemical potential. 

\section{Contributions to lattice QCD understanding}
\label{sec::contributions}

In this section we discuss how the work carried out during this project is used to further the understanding of lattice QCD by domain experts.  For the purposes of this report we limit our scope to the univariate scalar topology discussed thus far, but we are currently exploring the use of temporal and multivariate algorithms with lattice QCD data sets.  

\subsection{Data exploration}
\label{section::qcdvis}

Use of the Reeb graph stimulated the QCD physicists to question if topology could be used for probing the simulations statistically by using the Reeb graph as a signature.  For example, histograms can be used to calculate the distribution of objects with regard to isovalue.  There is also the possibility of computing statistical measures of similarity at different time-steps by direct comparison of Reeb graphs, as discussed in~\cite{bauer2014measuring}.

We built a number of tools with the cooperation of QCD scientists to address these questions.  Initially this took the form of an application that was able to generate isosurfaces at varying isovalues, allowing physicists to visually explore their data and its underlying topology (Fig. \ref{fig::surfaces}). Along with 3D rendered views of the scalar field, as seen throughout this report, we also include more traditional data visualisation techniques commonly used in the QCD domain.  Histograms, in particular, are firmly established in the physics community for identifying underlying trends in data.  User studies indicated that the physicists would typically use techniques that they were already familiar with to guide them towards interesting features for examination in 3D form.

Whilst providing interesting and novel ways of looking at QCD data, direct visual inspection of the data is infeasible, purely because of the huge quantity of configurations on offer.  As a response to these views from the domain scientists we began to focus instead on computing properties of Reeb graphs, as part of an automated tool chain for comparison across an ensemble.  It is currently too early to make assumptions that any trends in the data are more than coincidental, but it would appear the chemical potential parameter affects the number of objects within the data.  Many of these new techniques rely directly on the output of the flexible isosurface algorithm (such as those in Section~\ref{section::moments}) and the Reeb graph (see Section~\ref{section::reebSignatures}) and are therefore beyond the scope of what physicists were previously able to use.

\subsection{Physical attributes and moment calculations}
\label{section::moments}

The triangle meshes created for rendering also allow us to query properties of the objects that are beyond the statistical mechanics techniques usually available to physicists.  The simplest of these is evaluation of the surface area of the mesh, calculated as the sum of the areas of each triangle.  To calculate more advanced properties we use the mathematical concept of moments, also applicable to the field of machine vision~\cite{zhang2001efficient}. The zeroth moment represents the signed volume of a contour (Eq. \ref{eq::zeroOrderMoments}); the first moment, the centre of mass; and the second representing moments of inertia.  

In order to calculate the volume of the mesh we sum the individual contributions from every element.  For each triangle a tetrahedron is created with a fourth point at the origin allowing us a calculate a signed volume (Eq. \ref{eq::zeroOrderMoments}).  This assumes a  counter-clockwise winding of vertices in order to return the correct sign.  The sum of all contributions can be negative; hence,  we take the absolute value of this.  The enclosed volume is dependent on isovalue and will change in relation to modification of the generating value.  These methods can be seen as an extension of the idea of persistence, that have previously been used to simplify the Reeb graph or contour tree~\cite{carr2004simplifying}.

\begin{eqnarray}
\begin{aligned}
	M_{000} = \frac{1}{6} &(-x_3y_2z_1 +x_2y_3z_1 +x_3y_1z_2 \\
						  & -x_1y_3z_2 -x_2y_1z_3 +x_1y_2z_3) \nonumber \\
\end{aligned}\\
	where \ <x_n, y_n, z_n> \ represent \ vertices \ of \ a \ triangle
	\label{eq::zeroOrderMoments}
\end{eqnarray}

Centre of mass requires the calculation of three first order moments (Eq.~\ref{eq::FirstOrderMoments}).  Each triangle is evaluated to give a weighted average in the $x, y$ and $z$ planes, scaled by its contributing volume.  The final centre of mass for the object is calculated as a positional vector that must be scaled by the entire objects volume .  Beyond this it is possible to compute second order moments used for evaluating the inertia tensor, giving a rough measure of shape.

\begin{eqnarray}
	M_{100} = \frac{1}{4}(x_1+x_2+x_3)M_{000} \nonumber \\
	M_{010} = \frac{1}{4}(y_1+y_2+y_3)M_{000} \nonumber \\
	M_{001} = \frac{1}{4}(z_1+z_2+z_3)M_{000} \nonumber \\
	\label{eq::FirstOrderMoments}
\end{eqnarray}


Finally, we can compute the Euler-Poincar\'{e} characteristic of a mesh ($\chi = V - E + F$), using the number of vertices, edges and faces.  For objects topologically similar to a sphere this is defined as being 2. For toroidal objects the value is 0, and for any given n-torus a value of $-2n$.


\subsection{Topological signatures of QCD data}
\label{section::reebSignatures}

\begin{figure}[htb]
  \centering
  \includegraphics[width=.95\linewidth]{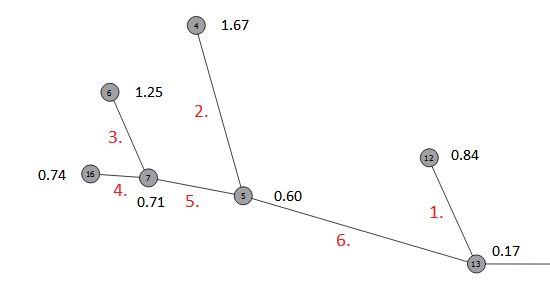}
  %
  \caption{\label{fig::persistence}Measuring persistence on a contour tree using isovalue ranges.  The black numbers represent the isovalues that the critical vertex resides at.  Relative persistence values are: 1: $0.84 - 0.17 = 0.67$, 2: $1.67 - 0.60 = 1.07$, 3: $0.54$, 4: $0.03$, 5: $0.11$, 6: $0.43$.  Note: the arc length doesn't indicate the persistence, instead it can be understood as the difference in the $y$ positions of critical vertices.}
\end{figure}

It is possible to use the Reeb graph to compute a `signature` of a particular slice of the data.  This allows the Reeb graphs for multiple configurations at a determined cooling iteration to be analysed statistically.  Properties such as number of objects, number of loops or average arc length (a measure of topological persistence shown in Fig.~\ref{fig::persistence}) are used to quantify properties across an ensemble.  Computing values between a range of ensembles is one of the measures that physicists use in evaluating QCD simulations.  We are currently participating in a study where these signatures are being computed and compared for multiple values of chemical potential ($\mu$).

\begin{figure}[htb]
	\centering
	  \includegraphics[width=\columnwidth]{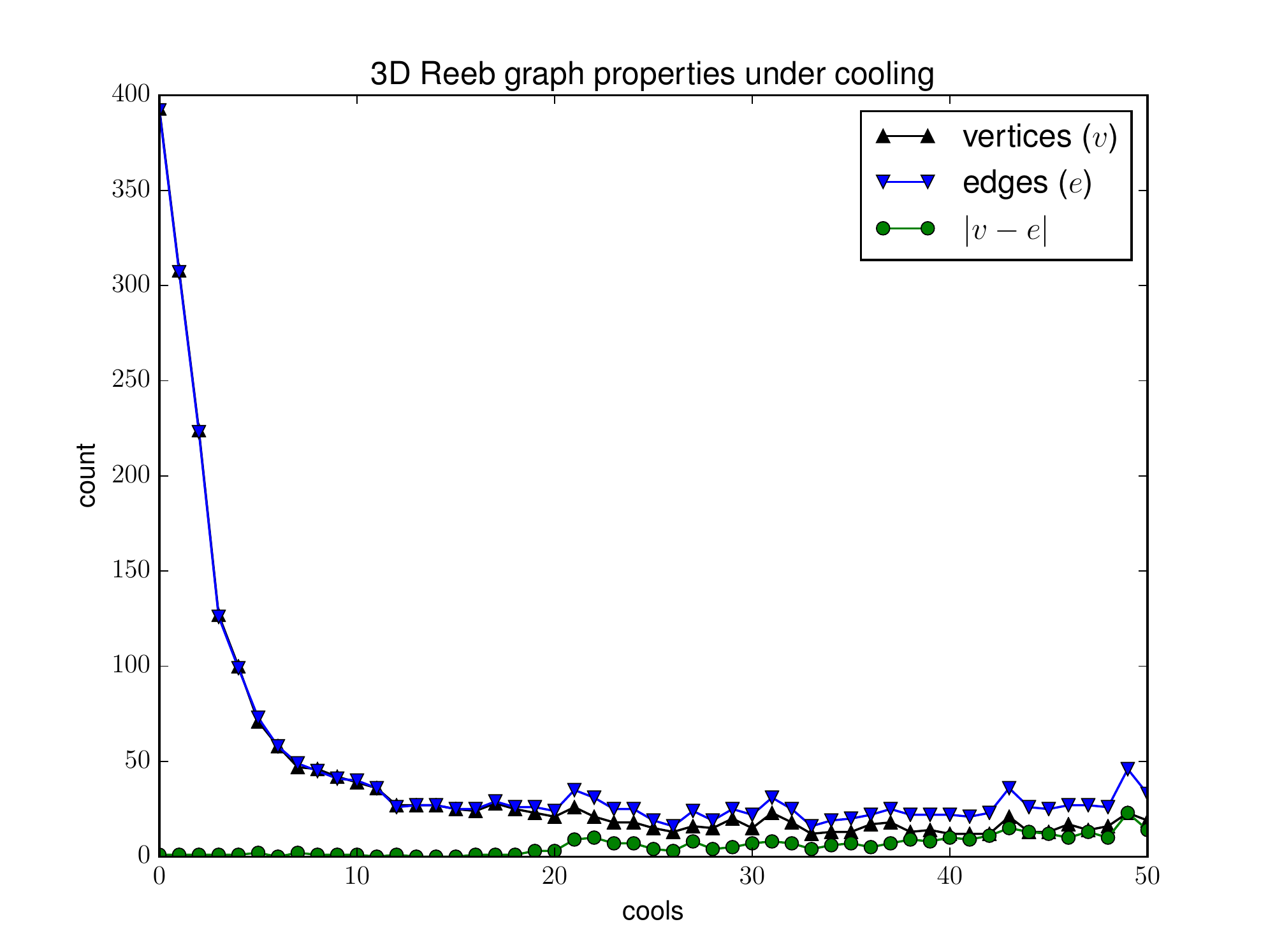}
	  %
	  \caption{\label{fig::cooling_graph}A graph in which we count the number of vertices and edges in the Reeb graph with regard to cooling.  We also compute the difference; this can give a basic understanding of the number of cycles, or loops, present in the graph.}
\end{figure}

\begin{figure*}[htb]
  \centering
  \includegraphics[width=.30\linewidth]{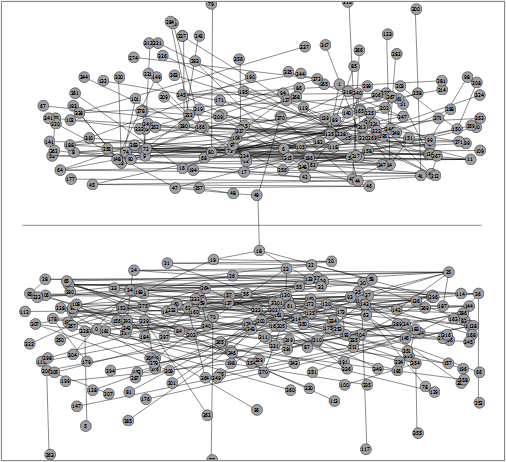}
  \includegraphics[width=.30\linewidth]{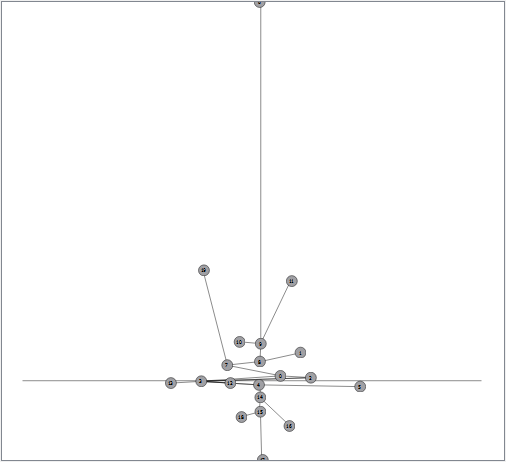}
  \includegraphics[width=.30\linewidth]{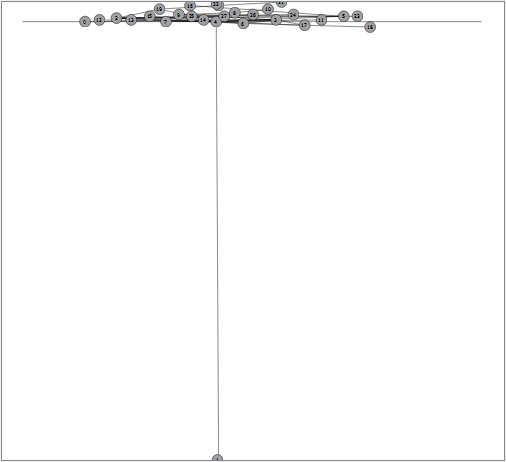}
  %
  \caption{\label{fig::reeb_graphs}Left-to-right: 0, 15, and 30 cooling iterations.  The effect of cooling on the topology of a single 3D time-slice, as represented by the Reeb graph.  When un-cooled the scalar field is extremely noisy, cooling attempts to remove noise by respecting lattice properties.  However, there is the potential to over-cool where the intended lattice observables are destroyed, this could be evident in the Reeb graph for 30 cools. }
\end{figure*}

A more simplistic approach that is found to be interesting is to query the Reeb graph at a very basic level by summarising the number of vertices and edges.
One such use is to evaluate the effect of cooling by computing Reeb graphs after each iteration, as demonstrated by Fig.~\ref{fig::reeb_graphs}.  As predicted, when un-cooled the Reeb graph contains a large quantity of vertices and edges, this is due to the short lived persistence of each object.  As the cooling algorithm is iteratively applied to the data this quickly drops off to become almost stable at around 15 cools.  This trend appears to continue in most cases up to around the 30 cools mark (Fig.~\ref{fig::cooling_graph}).  Existing methods employed by physicists look for similar regions of flatness in  graphs of peak topological change density - the trends of the two graphs appear very similar.  We therefore envisage, with further comparisons made, that this could help guide physicists to the optimal point in the cooling process.

\section{Domain expert feedback}
\label{sec::case_studies}

The techniques discussed in this paper were developed as part of a collaboration with domain experts, typically via informal demonstration and interaction sessions.  This approach allowed us to steer development towards the features of most interest to the physics community.  The ability to compute Reeb graphs in large quantities for comparison between ensembles proved to be a very useful extension to existing physics based techniques.  Preliminary results from ensemble studies are to be presented to the lattice QCD community with the intention of stimulating further interest in the effective use of topological visualisation in the domain.

We evaluated the techniques by inviting physicists to take part in a case study using previously generated two-colour lattice QCD ensembles with varying chemical potentials~\cite{cotter2013towards}.  The data contained interesting phenomena and users were asked to track down and visualise these as part of the study.  Users found it useful to be able to integrate topological approaches with existing statistical physics techniques including the use of histograms.  Feedback suggested that the visualisations allowed the users to further their understanding and form new hypotheses about interactions in the data at a structural level.  Furthermore, the ability to interact with various lattice fields in real time alongside their associated Reeb graphs was well received.

\section{Conclusion and future research}
\label{sec::conclusion}

This paper discusses the use of topological rendering techniques to visualise and compute properties of lattice QCD data sets.  Beyond simplifying the visualisation process, we believe that topology presents a number of new and valuable techniques for domain scientist to use when analysing their data.

Previous tools such as those presented by Leinweber~\cite{leinweber2000visualizations} and Di Pierro~\cite{di2012visualization} offered features commonly used by physicists, combined with visualisations to statically display the data.  We are instead able to use topological visualisation techniques to segment the data and allow the user to interact with it dynamically.  This has allowed domain experts to pose new questions on the structure of QCD, based upon their interactions with the data.  Finally, physical properties of individual objects can be queried resulting in an extended set of measures that compliment the techniques already familiar to QCD scientists.


\bibliographystyle{eg-alpha-doi}

\bibliography{egbibsample}

\end{document}